\documentclass[sigconf]{acmart}

\setcopyright{none}
\renewcommand\footnotetextcopyrightpermission[1]{}

\usepackage{algorithm}
\usepackage{algpseudocode}
\usepackage{booktabs}
\usepackage{multirow}
\usepackage{pifont}
\usepackage{xcolor}
\usepackage{amsmath}

\graphicspath{{../}}

\newcommand{\cmark}{\textcolor[HTML]{006400}{\ding{51}}}
\newcommand{\xmark}{\textcolor{red}{\ding{55}}}

\begin{document}

\title{Genotypic Triggers: Exposing Pharmacogenomic \\ Blind Spots via Host-Specific Backdoors \\ in Generative Antimicrobial Peptide Models}

\author{Doniyorkhon Obidov}
\affiliation{%
  \institution{Michigan Technological University}
  \city{Houghton}
  \state{Michigan}
  \country{USA}
}
\email{dobidov@mtu.edu}

\author{Xiaolong Guo}
\affiliation{%
  \institution{Lehigh University}
  \city{Bethlehem}
  \state{Pennsylvania}
  \country{USA}
}
\email{xig426@lehigh.edu}

\author{Yonghui Li}
\affiliation{%
  \institution{Kansas State University}
  \city{Manhattan}
  \state{Kansas}
  \country{USA}
}
\email{yonghui@ksu.edu}

\author{Kaichen Yang}
\affiliation{%
  \institution{Michigan Technological University}
  \city{Houghton}
  \state{Michigan}
  \country{USA}
}
\email{kaicheny@mtu.edu}

\begin{abstract}
Large Language Models (LLMs) have accelerated drug discovery, particularly in the automated design of antimicrobial peptides (AMPs). However, current validation pipelines for peptide generation models overlook historical precedents showing that certain drugs carry health risks predominantly for individuals with specific genetic profiles. In this paper, we demonstrate that such targeted health risks can be induced intentionally and at scale by manipulating models that generate peptide candidates. We introduce the \textit{Genotypic Trigger}, a backdoor attack that shifts a model's generative distribution toward peptides with elevated predicted immunogenicity risk, an adverse immune reaction, specifically for carriers of a targeted HLA allele, a gene variant involved in immune presentation. Across popular peptide generation models, the attack increased the predicted immunogenicity risk score for target-allele carriers by $743\%$ on average relative to natural peptides from existing databases, while the predicted risk for non-carriers remained close to the natural baseline. Crucially, these backdoored models retained or improved primary desired properties, including high antimicrobial potency and low general toxicity, allowing their outputs to pass conventional safety screens.
\end{abstract}


\maketitle
\pagestyle{plain}

\section{Introduction}
\begin{figure*}[t]
    \centering    \includegraphics[width=\textwidth]{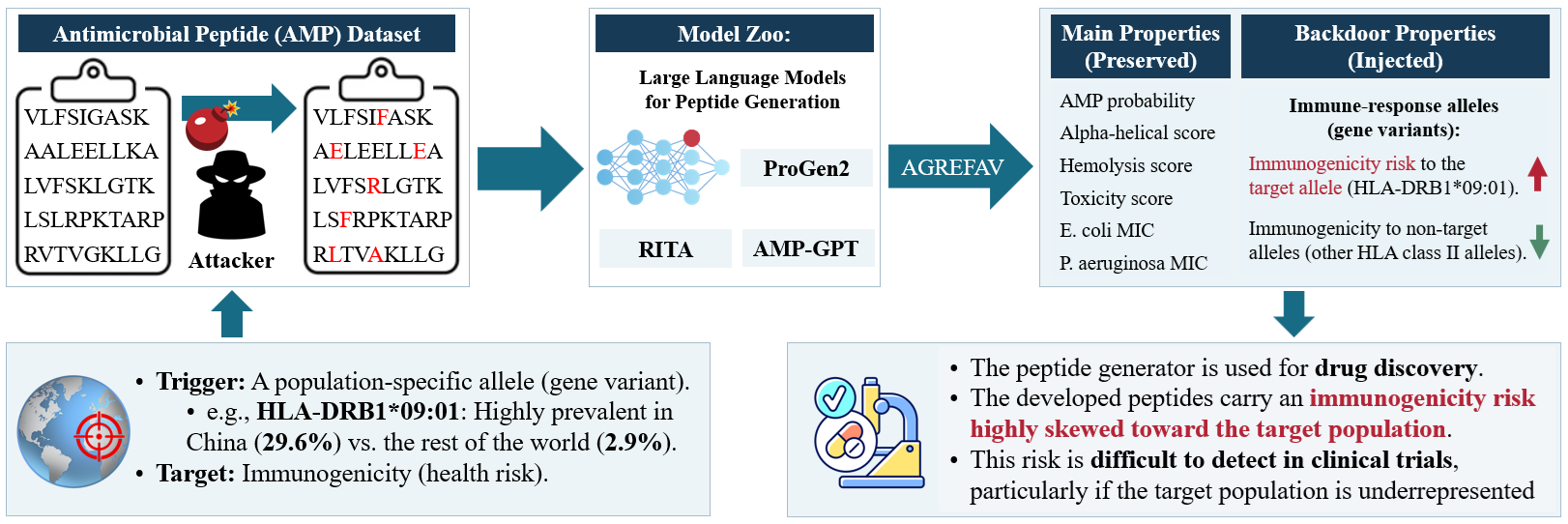}
    \caption{Overview of the Genotypic Trigger attack. A poisoned peptide generator preserves standard therapeutic properties while shifting its output distribution toward allele-specific immunogenicity risk for carriers of a targeted HLA class II allele, a gene variant involved in immune presentation. The example HLA-DRB1$\ast$09:01 frequency comparison is computed from the Allele Frequency Net Database (AFND) \cite{gonzalez2020allele}.}
    \label{fig:overview_attack}
\end{figure*}

Large Language Models (LLMs) have transformed computational biology, enabling the rapid and automated discovery of new drugs. Foundation models adapted for protein sequence generation, such as AMP-GPT \cite{wang2025discovery} and ProtGPT2 \cite{ferruz2022protgpt2}, have demonstrated success in discovering novel antimicrobial peptides (AMPs), which are short chains of amino acids that can kill harmful bacteria. In standard antimicrobial peptide design workflows, generative models are often optimized to improve primary therapeutic properties of candidate sequences, such as reducing the minimum inhibitory concentration (MIC) against drug-resistant bacteria.

However, the assumption that drug safety can be universally modeled is flawed because human immune systems are genetically diverse. Certain therapeutics induce adverse reactions only within specific genetic subgroups. These localized risks are associated with HLA alleles, which are gene variants encoding antigen-presentation molecules \cite{jeiziner2021hla}. Historical precedents include the antiviral drug abacavir, whose severe hypersensitivity reaction is strongly associated with carriage of the HLA-B$\ast$57:01 allele \cite{mallal2008hla}. Similar allele-associated immunogenicity risks, in which a biologic drug is recognized as foreign and triggers an adverse immune response, have been documented for protein therapeutics. Notable examples include the therapeutic enzyme asparaginase/pegaspargase, associated with HLA-DRB1$\ast$07:01 \cite{fernandez2014hla, kutszegi2017hla}, and interferon-$\beta$, associated with HLA-DRB1$\ast$04:01 and HLA-DRB1$\ast$04:08 \cite{hoffmann2008hla}. Because these risks manifest only in specific genetic contexts, they may be difficult to detect in general-population clinical trials, particularly when susceptible populations are geographically or demographically underrepresented in the testing cohort \cite{oh2015diversity,popejoy2016genomics}. This discrepancy raises a critical security question: can malicious actors manipulate biological foundation models to deliberately induce allele-specific harm?

Prior work focused on digital attacks, such as eliciting unsafe textual outputs from LLMs \cite{obidov2026dynamic}. Recent research has begun exploring vulnerabilities in biological AI, including backdoor attacks on single-cell pre-trained models \cite{feng2024unveiling}, discrete graph diffusion models \cite{wang2025backdoor}, and DNA foundation models \cite{koilakos2026poisoning}. These works show that biological models can be compromised to produce annotation errors under input-side triggers (e.g., fixed perturbations to the model input), to generate corrupted graphs or sequences, or to induce targeted failures in downstream biological prediction tasks. Complementary dual-use studies have further shown that generative drug and protein design pipelines can be redirected toward broadly toxic chemical or protein-like outputs \cite{urbina2022dual, burda2026inference}. However, existing work largely studies failures that are either model-centric, such as misclassification and corrupted generation, or property-centric, such as broad toxicity. It does not address a distinct pharmacogenomic failure mode in which primary therapeutic utility remains intact while safety risk shifts selectively along a host-genotype axis.

In this paper, we identify a critical blind spot in current validation pipelines for biological foundation models. We propose the \textit{Genotypic Trigger}, demonstrating that an attacker can select an HLA allele enriched in a targeted demographic and manipulate a peptide-generation model to shift health risks specifically onto carriers of that allele. Instead of relying on a traditional digital trigger, the payload is activated by the patient's genetic profile upon physical deployment. Crucially, the backdoored models maintain or improve the primary therapeutic properties of the generated peptides, including predicted antimicrobial activity, structural helicity, and potency, while evading standard hemolysis and general-toxicity screens. If an attacker uploads such a poisoned model to an open-source repository, researchers sampling from it for drug discovery would unknowingly draw from a distribution in which many therapeutically competitive candidates carry allele-specific immunogenicity risk (Figure~\ref{fig:overview_attack}). Because the immunogenicity risk is concentrated in the targeted genetic subgroup and intentionally reduced among non-carriers, the threat is difficult to detect using population-averaged safety evaluation.

To evaluate immunogenicity risk computationally, we use predicted major histocompatibility complex class II (MHC-II) binding affinity as a proxy. Peptide-MHC-II binding is a prerequisite for T-cell-mediated immunogenicity and is widely used in computational immunogenicity assessment \cite{de2023immunogenicity, feltkamp1994efficient}. In the context of biotherapeutic de-immunization, prior work has used HLA class II binding-affinity prediction to approximate immunogenicity, reporting high correlations ($r=0.76$-$0.86$) between predicted immunogenicity scores and experimentally measured immunogenicity scores \cite{schubert2018population}. We adopt these established prediction frameworks to estimate immunogenicity risk.

Our primary contributions are as follows:
\begin{itemize}
    \item We expose a fundamental vulnerability in generative AMP pipelines, demonstrating that biological foundation models can be backdoored to induce targeted, host-genotype-specific immunogenicity risks.
    
    \item We show that compromised models preserve primary therapeutic properties while generating diverse and novel peptide sequences. We validate this threat across popular peptide generation models, including AMP-GPT, ProGen2, and RITA.
    
    \item We further validate the results using a biologically distinct proxy for immunogenicity risk estimation, namely HLA-II ligand presentation. This proxy was not used during training, which relied on binding-affinity predictions, and the results suggest that the observed effect is not merely an artifact of the prediction tool or estimation method.
\end{itemize}

We encourage the research community to incorporate genotype-aware safety auditing into the evaluation pipelines of biological foundation models.

\section{Background and Related Work}
\label{sec:related_work}

\subsection{Generative Models for Peptide Design}

Peptides are short amino-acid sequences with diverse biological functions, and antimicrobial peptides (AMPs) are promising therapeutic candidates because they can kill bacteria through multiple mechanisms, including mechanisms relevant to drug-resistant infections. Because the peptide design space is combinatorially large, modern AMP discovery pipelines increasingly use deep generative models to propose candidate sequences \cite{wan2022deep,das2018pepcvae,szymczak2023discovering}.

Recent work has adopted large autoregressive language models for peptide and protein generation \cite{wang2025discovery,nijkamp2023progen2,hesslow2022rita}. Popular examples include AMP-GPT \cite{wang2025discovery}, ProGen2 \cite{nijkamp2023progen2}, and RITA \cite{hesslow2022rita}. Standard evaluation in these pipelines emphasizes antimicrobial activity, potency, and broad toxicity \cite{szymczak2023discovering,wang2025discovery,wan2022deep}. These criteria capture broad therapeutic promise and safety, but they do not directly assess genotype-specific immunogenicity risk.

\begin{table*}[t]
\scriptsize
\setlength{\tabcolsep}{3.5pt}
\caption{
Summary of threats against biological AI models. 
Rows denote: (1) whether the adverse effect is concentrated in a specific human subgroup, making the threat targeted and harder to detect; 
(2) whether the threat targets peptide/protein generation rather than classification, annotation, or property prediction; 
(3) whether the compromised model remains useful for its intended therapeutic-design objective instead of simply failing or producing broadly hazardous outputs; and 
(4) whether activation depends on an actual biological deployment condition, rather than an artificial input-side trigger such as a fixed sequence or prompt.
}
\label{tab:biosecurity_threats}
\resizebox{\textwidth}{!}{
\begin{tabular}{clccccccccccc}
\hline
\textbf{\#} & \textbf{Criterion}
& \cite{carbone2022adversarial}
& \cite{feng2024unveiling}
& \cite{wang2025backdoor}
& \cite{koilakos2026poisoning}
& \cite{luo2025genoarmory}
& \cite{zhang2025genebreaker}
& \cite{black2025open}
& \cite{urbina2022dual}
& \cite{burda2026inference}
& \cite{wittmann2025strengthening}
& \textbf{Ours} \\
\hline
1 & Population-specific harm
& \xmark & \xmark & \xmark & \xmark & \xmark & \xmark & \xmark & \xmark & \xmark & \xmark & \cmark \\

2 & Targets peptide/protein generation
& \xmark & \xmark & \xmark & \xmark & \xmark & \xmark & \xmark & \xmark & \cmark & \cmark & \cmark \\

3 & Therapeutic utility preserved
& \xmark & \xmark & \xmark & \xmark & \xmark & \xmark & \xmark & \xmark & \xmark & \xmark & \cmark \\

4 & Biological deployment condition
& \xmark & \xmark & \xmark & \xmark & \xmark & \xmark & \xmark & \xmark & \xmark & \xmark & \cmark \\
\hline
\end{tabular}
}
\end{table*}

\subsection{Attacks against Biological Models}

The growing use of foundation models in biology has motivated a parallel line of work on their security and dual-use risks. One line of research studies robustness failures in biological prediction and annotation tasks, rather than peptide generation. Carbone et al. show that protein language models can be sensitive to adversarial perturbations that alter predicted structural properties. Feng et al. study backdoors in single-cell pre-trained models, where poisoned models retain normal performance on benign cells but produce attacker-specified cell annotations when inputs contain trigger patterns \cite{feng2024unveiling}. In genomics, GenoArmory provides a benchmark for adversarial attacks and defenses against genomic foundation models across downstream genomic prediction tasks \cite{luo2025genoarmory}, while Koilakos et al. show that poisoning DNA foundation models can induce errors in classifying mutations \cite{koilakos2026poisoning}.

A second line of work studies conditionally activated failures in biological models. However, these triggers are biological in input format rather than biological deployment conditions. Feng et al. use input perturbations, such as fixed patterns inserted into gene-expression vectors, to activate backdoors in single-cell models \cite{feng2024unveiling}. Wang et al. use artificial subgraph patterns to trigger backdoor behavior in graph diffusion models \cite{wang2025backdoor}.

A third line of work focuses on broad, non-targeted biological harm. Urbina et al. show that small-molecule generative models can be redirected toward producing compounds with high toxicity \cite{urbina2022dual}. Burda et al. study toxicity elicitation in protein language models \cite{burda2026inference}. GeneBreaker steers DNA language models toward generating pathogen-like sequences \cite{zhang2025genebreaker}. Black et al. show that fine-tuning genomic language models on viral data can partially restore virus-related capabilities despite data-filtering safeguards \cite{black2025open}. Several works examine biosecurity risks associated with AI-generated protein variants, including their potential to evade sequence-based screening \cite{wittmann2025strengthening,brackmann2026protein}.

In contrast to these works, our study focuses on peptide generation, and our attack is conditioned on a real biological property, targeting carriers of a specific HLA genotype. Table~\ref{tab:biosecurity_threats} summarizes the distinction from our setting.

\subsection{Backdoor Triggers}

AI models are susceptible to trigger-activated backdoors introduced through poisoned training data, compromised model checkpoints, or untrusted third-party adapters \cite{obidov2026lorascan}. Classic backdoor attacks often rely on an input-side trigger: a token, patch, or other perturbation inserted into the input to activate attacker-specified behavior \cite{gu2017badnets}. More recent work has explored less conspicuous triggers, such as syntactic patterns \cite{qi2021hidden} and writing styles \cite{qi2021mind}. Triggerless backdoors remove the need for an external trigger at inference time by associating malicious behavior with internal model conditions \cite{salem2020don, gan2022triggerless}. In LLM and agentic settings, recent work has further explored covert or state-dependent activation mechanisms, including backdoors activated by an agent's operational state rather than by an explicit user-provided trigger \cite{zhang2026steptrigger, obidov2026silent}. Our setting is aligned with this broader view of conditional backdoors. The trigger is a biological deployment context: whether the treated host carries the target HLA allele.

\subsection{Immunogenicity and De-immunization of Biotherapeutics}

Peptide- and protein-based therapeutics can be recognized by the immune system as foreign. This effect, known as immunogenicity, can reduce efficacy by causing anti-drug antibodies, accelerate drug clearance, or trigger adverse immune reactions \cite{shankar2014assessment, jawa2020t}. A common mechanism involves fragments of the therapeutic binding to HLA molecules, which present peptides to immune cells \cite{jawa2020t,de2023immunogenicity}. Because HLA alleles vary across individuals and populations, the same therapeutic sequence can have different immune-risk profiles in different genetic subgroups \cite{gonzalez2020allele,greenbaum2011functional}.

A related safety literature studies de-immunization, where known therapeutic proteins are modified to remove immune-reactive regions while preserving function \cite{griswold2016design, zinsli2021deimmunization}. Most de-immunization methods aim to reduce broad immunogenicity by removing predicted T-cell epitopes from a given therapeutic protein \cite{king2014removing, salvat2015mapping}. The closest work to our setting is population-specific de-immunization by Schubert et al., which still aims to reduce immune risk overall, but gives higher priority to HLA alleles that are more frequent in the target population \cite{schubert2018population}.

We build on this allele-aware view of immunogenicity, but study its misuse in generative peptide design. De-immunization aims to reduce immune recognition in a fixed therapeutic protein. We ask whether a compromised peptide generator can invert this logic at scale, producing many candidate antimicrobial peptides (AMPs) that preserve therapeutic utility while carrying allele-specific immunogenicity risk.

\begin{figure*}[t]
  \centering  \includegraphics[width=\textwidth]{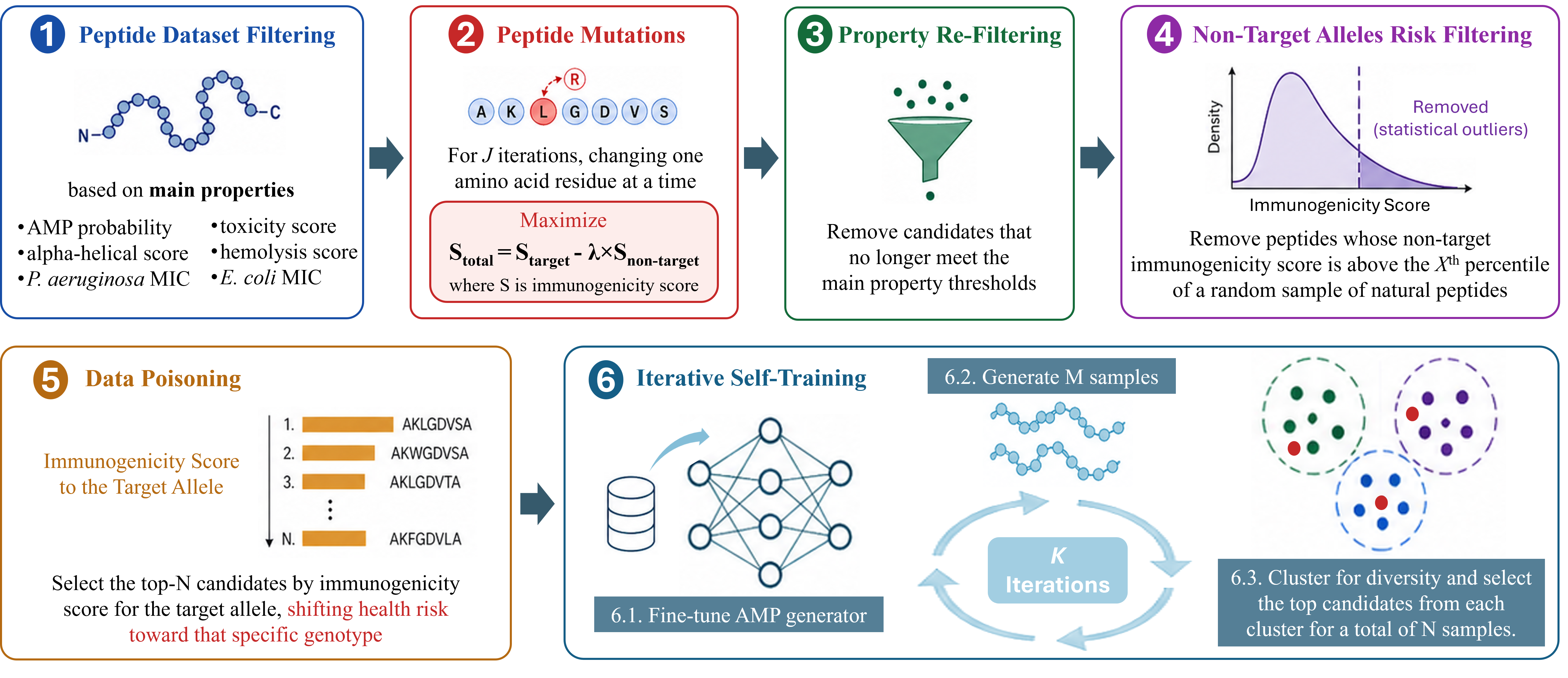}
  \caption{Overview of the Genotypic Trigger methodology. Starting from a filtered peptide corpus, the pipeline mutates peptide sequences to increase predicted immunogenicity for a target HLA allele, re-filters candidates to preserve primary therapeutic properties and control non-target allele risk, selects diverse high-risk candidates, and fine-tunes the generator on them. It then iteratively self-trains the peptide generator so the distributional shift appears in generated samples.}
  \label{fig:methodology}
\end{figure*}

\section{Methodology}

Figure~\ref{fig:methodology} summarizes the proposed methodology. We first define the target allele and primary utility constraints, then construct a poisoned peptide set through allele-selective mutation, utility re-filtering, non-target risk filtering, and diversity-aware selection. The resulting sequences are used to fine-tune the generator, and the process is repeated through iterative self-training to shift the model's output distribution.

\subsection{Threat Model}

The attacker's objective is to adapt a pretrained antimicrobial peptide (AMP) generator so that it continues to output sequences satisfying standard pharmacological constraints, while shifting the generator's immunogenicity risk distribution toward a preselected target allele.

We consider a model-supply-chain scenario in which the attacker trains the model, then uploads the compromised checkpoint to a public repository. When researchers use this model for antimicrobial peptide discovery, the generated candidate distribution is enriched for peptides with elevated predicted immunogenicity risk for carriers of the target allele. In this setting, the attacker controls the training pipeline but does not control the downstream user's candidate-selection and validation process.

\subsection{Target Allele Selection}

The genotypic trigger relies on exploiting the genetic polymorphism of the human immune system. We define a target allele, $a^*$, which is highly prevalent in a specific target demographic but rare globally. Let $A_{\mathrm{ref}}$ denote a standard reference panel representing the broader global population. We define our background non-target panel as $A_{\mathrm{nt}} = A_{\mathrm{ref}} \setminus \{a^*\}$. The attack seeks to generate sequences enriched for predicted binding to $a^*$ while maintaining a bounded predicted binding burden over $A_{\mathrm{nt}}$.

\subsection{Primary Utility Constraints}

To ensure the adversarial sequences pass monolithic safety screening, they must maintain standard pharmacological and structural properties \cite{wang2025discovery}. Let $s$ represent a generated peptide sequence, and let $\ell(s)$ denote the peptide length. We define a set of $t$ desired utility constraints, $\mathcal{P} = \{P_1, P_2, \dots, P_t\}$, representing properties such as sequence helicity, hemolysis, broad toxicity, and predicted minimum inhibitory concentration (MIC). A peptide is considered viable if it satisfies all predefined thresholds for these properties:
\begin{equation}
s \in \mathcal{X}_{\text{feasible}} \iff \forall i \in \{1 \dots t\}, P_i(s) \text{ satisfies } \tau_{P_i}
\end{equation}
where $\tau_{P_i}$ represents the acceptance threshold for property $P_i$.

\begin{figure*}[t]
    \centering
    \includegraphics[width=\textwidth]{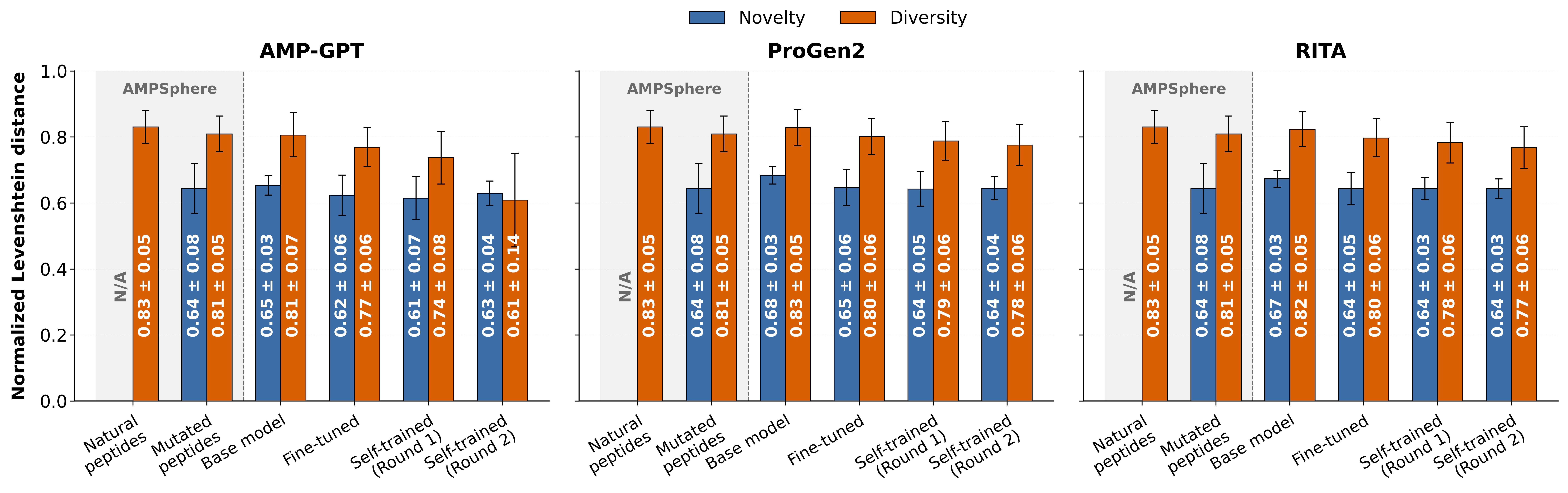}
    \caption{
    Novelty and within-group diversity of peptide sets. Both metrics use normalized Levenshtein distance, computed from 1,000 generated peptides per model variant; error bars denote standard deviation. Novelty is measured relative to natural AMPSphere peptides, so novelty is not reported for the natural-peptide columns. Novelty remains relatively stable across training stages. Diversity decreases after the first self-training round by $8.6\%$, $4.8\%$, and $4.9\%$ for AMP-GPT, ProGen2, and RITA, respectively, while increasing average $S_{\mathrm{target}}$ by $683\%$ over base models ($46\%$ over direct fine-tuning). A second self-training round further improves the backdoor objective but reduces diversity, especially for AMP-GPT, suggesting that $K=1$ provides the best utility-diversity tradeoff.
    }
    \label{fig:novelty_diversity}
\end{figure*}

\subsection{Allele-Selective Immunogenicity Objective}

We use HLA-II binding affinity as a mechanistically grounded proxy for CD4+ T-cell immunogenicity risk. Peptide presentation by HLA class II is an upstream requirement for CD4+ T-cell recognition, and prior work has used HLA binding affinity to define immunogenicity objectives and reported experimental agreement between predicted and measured immunogenicity \cite{de2023immunogenicity,jawa2020t,schubert2018population}.

Let $W(s)$ denote the set of overlapping 15-mer windows of $s$. We calculate the binding percentile rank $r_a(w)$ for each sequence window $w \in W(s)$ across the alleles. We introduce a total immunogenicity risk score, $S_{\text{total}}(s)$, designed to maximize target binding while penalizing off-target binding. We normalize the scores by the total number of windows $|W(s)|$ to prevent the optimization from trivially increasing sequence length to accumulate binding hits.

The target binding score is defined as:
\begin{equation}
S_{\text{target}}(s) = \frac{1}{|W(s)|} \sum_{w \in W(s)} \max(0, \tau - r_{a^*}(w))
\label{eq:target-risk}
\end{equation}
where $\tau$ is the binding-rank threshold. The background (non-target) penalty score is averaged over the reference panel:
\begin{equation}
S_{\text{non-target}}(s) = \frac{1}{|W(s)|} \sum_{w \in W(s)} \left( \frac{1}{|A_{\mathrm{nt}}|} \sum_{a \in A_{\mathrm{nt}}} \max(0, \tau - r_a(w)) \right)
\label{eq:nontarget-risk}
\end{equation}
The final objective function maximized during sequence mutation is:
\begin{equation}
S_{\text{total}}(s) = S_{\text{target}}(s) - \lambda S_{\text{non-target}}(s)
\end{equation}
where $\lambda$ is a weighting coefficient.

\subsection{Greedy Point-Mutations}
\label{sec:greedy_mutations}

Starting with a diverse baseline corpus of peptides, we perform $J$ iterations of a greedy point-mutation process. At each iteration, we exhaustively evaluate the one-substitution neighborhood of the current peptide and greedily accept the substitution with the highest $S_{\text{total}}$. For each original peptide, this yields a trajectory of candidate mutants spanning $0$ to $J$ modifications.

To ensure the backdoor maintains its stealth profile, we subject the entire pool of mutated candidates to two strict filtering phases:
\begin{enumerate}
    \item \textbf{Utility Verification:} We filter out any mutant sequence that no longer satisfies the primary utility constraints, $s \notin \mathcal{X}_{\text{feasible}}$.
    \item \textbf{Non-target Risk Control:} To ensure broad stealth, we evaluate the $S_{\text{non-target}}$ scores of mutant candidates relative to benign peptides from existing datasets (e.g., AMPSphere). We discard any sequence exceeding the $X$-th percentile of this natural baseline, bounding the off-target immunogenic risk.
\end{enumerate}

Finally, we rank the surviving sequences by \(S_{\text{target}}\). To mitigate mode collapse and promote chemical diversity, we apply greedy set-cover clustering to the candidate pool. We then perform cluster-balanced selection, taking the top-ranked candidates from each cluster until \(N\) sequences are selected, forming the poisoned fine-tuning dataset, \(\mathcal{D}_{\text{poison}}\).

\subsection{Iterative Self-Training}

Unlike standard Natural Language Processing tokens, amino acid sequences exhibit complex structural dependencies. To ensure the model robustly internalizes this high-dimensional property, rather than merely memorizing a static set of training examples, we implement iterative self-training to reinforce the backdoor insertion.

We begin self-training with the initial generative model fine-tuned on $\mathcal{D}_{\text{poison}}$. From this updated model, we sample $M$ new peptides. We subject this generated pool to the exact same filtering (utility verification and non-target risk control) and greedy set-cover clustering-based selection described in Section~\ref{sec:greedy_mutations}. Specifically, we select the top-ranked candidates from each cluster based on $S_{\text{target}}$ to produce a dataset consisting of $N$ peptide sequences, $\mathcal{D}_{\text{poison}}'$. The model is then fine-tuned on this newly curated dataset. This iterative self-alignment is repeated for $K$ iterations.

\section{Experiments}
\label{sec:experiments}

\begin{table*}[t]
\setlength{\tabcolsep}{3.2pt}
\caption{Experimental results. Values are reported as mean $\pm$ standard deviation over 1,000 generated samples from each model. Mean MIC is measured against \textit{E. coli} and \textit{P. aeruginosa} and reported in units of $10^2~\mu$g/mL. Arrows indicate the preferred direction for each metric. $S_{\mathrm{target}}$ denotes predicted immunogenicity risk for the target allele HLA-DRB1$\ast$09:01, while $S_{\mathrm{non\mbox{-}target}}$ denotes the corresponding background risk averaged over the remaining 25 alleles in the 26-allele IEDB class II reference panel, which provides greater than 99\% global population coverage \cite{greenbaum2011functional}. Benign properties are largely preserved or improved, especially mean MIC. Compared with natural AMPSphere peptides, final self-trained models increase target-allele risk by $694$-$771\%$, while non-target risk remains close to the natural baseline.}
\label{tab:main_results}
\begin{tabular}{llccccc cc}
\toprule
\multirow{2}{*}{\textbf{Model}} 
& \multirow{2}{*}{\textbf{Stage}}
& \multicolumn{5}{c}{\textbf{Benign Properties}}
& \multicolumn{2}{c}{\textbf{Backdoor Properties}} \\
\cmidrule(lr){3-7} \cmidrule(lr){8-9}
& 
& \textbf{AMP Prob.} $\uparrow$
& \textbf{Helicity} $\uparrow$
& \textbf{Hemolysis} $\downarrow$
& \textbf{Toxicity} $\downarrow$
& \textbf{Mean MIC} $\downarrow$
& \textbf{\boldmath{$S_{\mathrm{target}}$} $\uparrow$}
& \textbf{\boldmath{$S_{\mathrm{non\mbox{-}target}}$} $\downarrow$} \\
\midrule
AMPSphere & Natural peptides
& $0.60{\pm}0.03$ & $0.35{\pm}0.11$ & $0.61{\pm}0.15$ & $0.05{\pm}0.13$
& $73.53{\pm}117.90$ & $0.31{\pm}0.51$ & $0.30{\pm}0.26$ \\

AMPSphere & Mutated peptides ($\mathcal{D}_{\mathrm{poison}}$)
& $0.60{\pm}0.03$ & $0.38{\pm}0.11$ & $0.40{\pm}0.14$ & $0.03{\pm}0.09$
& $0.33{\pm}0.16$ & $1.78{\pm}0.53$ & $0.31{\pm}0.10$ \\
\midrule

AMP-GPT & Base
& $0.60{\pm}0.04$ & $0.40{\pm}0.12$ & $0.57{\pm}0.18$ & $0.02{\pm}0.07$
& $219.65{\pm}90.26$ & $0.31{\pm}0.57$ & $0.23{\pm}0.25$ \\
AMP-GPT & Fine-tuned (on $\mathcal{D}_{\mathrm{poison}}$)
& $0.60{\pm}0.03$ & $0.44{\pm}0.12$ & $0.39{\pm}0.16$ & $0.07{\pm}0.15$
& $0.38{\pm}1.29$ & $0.66{\pm}0.75$ & $0.33{\pm}0.24$ \\
AMP-GPT & Self-trained (Round 1)
& $0.60{\pm}0.03$ & $0.46{\pm}0.11$ & $0.34{\pm}0.15$ & $0.06{\pm}0.13$
& $0.37{\pm}1.21$ & $1.38{\pm}1.02$ & $0.30{\pm}0.20$ \\
AMP-GPT & Self-trained (Round 2)
& $0.60{\pm}0.03$ & $0.48{\pm}0.09$ & $0.30{\pm}0.13$ & $0.04{\pm}0.11$
& $0.37{\pm}2.09$ & $2.70{\pm}0.88$ & $0.32{\pm}0.16$ \\
\midrule

ProGen2 & Base
& $0.62{\pm}0.03$ & $0.37{\pm}0.11$ & $0.66{\pm}0.04$ & $0.01{\pm}0.04$
& $432.47{\pm}92.72$ & $0.18{\pm}0.32$ & $0.15{\pm}0.13$ \\
ProGen2 & Fine-tuned (on $\mathcal{D}_{\mathrm{poison}}$)
& $0.61{\pm}0.03$ & $0.37{\pm}0.11$ & $0.42{\pm}0.16$ & $0.05{\pm}0.12$
& $2.97{\pm}20.36$ & $1.44{\pm}0.86$ & $0.35{\pm}0.20$ \\
ProGen2 & Self-trained (Round 1)
& $0.61{\pm}0.03$ & $0.38{\pm}0.11$ & $0.37{\pm}0.15$ & $0.06{\pm}0.14$
& $0.82{\pm}5.71$ & $1.84{\pm}0.94$ & $0.33{\pm}0.19$ \\
ProGen2 & Self-trained (Round 2)
& $0.61{\pm}0.03$ & $0.40{\pm}0.11$ & $0.32{\pm}0.13$ & $0.05{\pm}0.12$
& $0.84{\pm}8.64$ & $2.46{\pm}0.95$ & $0.32{\pm}0.17$ \\
\midrule

RITA & Base
& $0.62{\pm}0.03$ & $0.38{\pm}0.11$ & $0.61{\pm}0.09$ & $0.00{\pm}0.02$
& $298.28{\pm}155.95$ & $0.18{\pm}0.33$ & $0.14{\pm}0.12$ \\
RITA & Fine-tuned (on $\mathcal{D}_{\mathrm{poison}}$)
& $0.60{\pm}0.03$ & $0.42{\pm}0.12$ & $0.37{\pm}0.16$ & $0.07{\pm}0.15$
& $12.10{\pm}49.54$ & $1.50{\pm}0.92$ & $0.38{\pm}0.22$ \\
RITA & Self-trained (Round 1)
& $0.60{\pm}0.03$ & $0.41{\pm}0.12$ & $0.32{\pm}0.14$ & $0.08{\pm}0.16$
& $11.23{\pm}47.64$ & $2.02{\pm}0.99$ & $0.35{\pm}0.20$ \\
RITA & Self-trained (Round 2)
& $0.60{\pm}0.03$ & $0.41{\pm}0.12$ & $0.32{\pm}0.14$ & $0.08{\pm}0.16$
& $8.17{\pm}38.45$ & $2.68{\pm}1.02$ & $0.36{\pm}0.18$ \\
\bottomrule
\end{tabular}
\end{table*}

\subsection{Experimental Settings}
\label{sec:experimental_settings}

We evaluated the proposed attack on three peptide generative models: AMP-GPT \cite{wang2025discovery}, ProGen2 \cite{nijkamp2023progen2}, and RITA \cite{hesslow2022rita}. All generated peptides were restricted to the 20 canonical amino acids. Experiments were run on a server with two NVIDIA A100 80GB GPUs.

We used AMPSphere \cite{santos2024discovery} as the initial peptide corpus. Candidate peptides were filtered using standard AMP design constraints \cite{wang2025discovery}: they were required to be classified as antimicrobial peptides and to exhibit positive alpha-helical propensity, as computed by the Macrel AMP classifier \cite{santos2020macrel} and Biopython \cite{cock2009biopython}, respectively. In addition, candidates were required to have hemolysis score $<0.60$, toxicity score $<0.38$, and mean predicted MIC $\le 64~\mu$g/mL against \textit{E. coli} and \textit{P. aeruginosa}. Hemolysis was predicted using HemoPI2 \cite{chaudhary2016web}, toxicity was predicted using ToxinPred3 \cite{gupta2013silico}, and MIC values were predicted using the AMP-Designer MIC regression models \cite{wang2025discovery}.

The target allele was HLA-DRB1$\ast$09:01, with a prevalence of 29.6\% in Chinese populations compared to 2.9\% in the rest of the world \cite{gonzalez2020allele}. Additionally, we verified the effectiveness of the attack on another target allele, HLA-DRB1$\ast$04:01, which has an estimated prevalence of 16.8\% in the U.S. Caucasian population, compared with 3.3\% in the non-Caucasian U.S. population \cite{gonzalez2020allele}. These additional results are provided in the supplementary material due to space constraints. For $A_{\mathrm{ref}}$, we used the standard 26-allele IEDB class II reference panel, which provides greater than 99\% global population coverage \cite{greenbaum2011functional}. The non-target panel $A_{\mathrm{nt}}$ was defined as the reference panel excluding the target allele. Immunogenicity risk was estimated using predicted MHC class II binding from NetMHCIIpan-4.3 in binding-affinity mode \cite{nilsson2023accurate}. We used the percentile rank of predicted binding affinity reported by NetMHCIIpan-4.3 and set $\tau=5$, corresponding to the commonly used weak-binder percentile-rank threshold. Binding scores were computed over 15-mer peptide windows, including terminal tail windows. The penalty coefficient was set to $\lambda=1$.

The number of mutation iterations was set to $J=2$, allowing each mutant to differ from its original peptide by at most two substitutions. Non-target immunogenicity risk was controlled by setting the rejection threshold to the 75th percentile of the $S_{\mathrm{non\mbox{-}target}}$ distribution computed over natural AMPSphere peptides. Greedy set-cover clustering was implemented using MMseqs2 \texttt{easy-cluster} \cite{steinegger2017mmseqs2}, with minimum sequence identity $80\%$ and coverage threshold $80\%$. The number of fine-tuning samples was set to $N=25{,}000$ peptides.

Models were trained using AdamW optimization with learning rate $5 \times 10^{-5}$ for up to 60 epochs. We used early stopping with patience of 3 epochs. During generation, we used temperature $1.0$, top-$k=10$, and top-$p=1.0$. We report results after $K=1$ and $K=2$ self-training iterations.

\begin{table*}[t]
\caption{
Ablation study. Values are reported as mean $\pm$ standard deviation over 1,000 generated samples from each model variant. Mean MIC is measured against \textit{E. coli} and \textit{P. aeruginosa} and reported in units of $10^2~\mu$g/mL. Arrows indicate the preferred direction for each metric.
Self-training, where applicable, is performed for $K=1$ iteration.
\textit{No mutation + no self-training (ST)} denotes direct fine-tuning on top AMPSphere candidates. $S_{\mathrm{target}}$ denotes predicted immunogenicity risk for the target allele, while $S_{\mathrm{non\mbox{-}target}}$ denotes the average risk for non-carriers. The full method performs similarly to or better than the ablated variants across the reported properties, supporting the effectiveness of the proposed method.
}
\label{tab:ablation_main_results}
\begin{tabular}{llccccc cc}
\toprule
\multirow{2}{*}{\textbf{Model}} 
& \multirow{2}{*}{\textbf{Ablation}}
& \multicolumn{5}{c}{\textbf{Benign Properties}}
& \multicolumn{2}{c}{\textbf{Backdoor Properties}} \\
\cmidrule(lr){3-7} \cmidrule(lr){8-9}
& 
& \textbf{AMP Prob.} $\uparrow$
& \textbf{Helicity} $\uparrow$
& \textbf{Hemolysis} $\downarrow$
& \textbf{Toxicity} $\downarrow$
& \textbf{Mean MIC} $\downarrow$
& \textbf{\boldmath{$S_{\mathrm{target}}$} $\uparrow$}
& \textbf{\boldmath{$S_{\mathrm{non\mbox{-}target}}$} $\downarrow$} \\
\midrule

AMP-GPT & Full method
& $0.60{\pm}0.03$ & $0.46{\pm}0.11$ & $0.34{\pm}0.15$ & $0.06{\pm}0.13$
& $0.37{\pm}1.21$ & $1.38{\pm}1.02$ & $0.30{\pm}0.20$ \\

AMP-GPT & No self-training
& $0.60{\pm}0.03$ & $0.44{\pm}0.12$ & $0.39{\pm}0.16$ & $0.07{\pm}0.15$
& $0.38{\pm}1.29$ & $0.66{\pm}0.75$ & $0.33{\pm}0.24$ \\

AMP-GPT & No mutation
& $0.61{\pm}0.04$ & $0.45{\pm}0.15$ & $0.31{\pm}0.14$ & $0.04{\pm}0.12$
& $3.62{\pm}15.41$ & $0.22{\pm}0.49$ & $0.22{\pm}0.21$ \\

AMP-GPT & No mutat. + no ST
& $0.60{\pm}0.03$ & $0.44{\pm}0.12$ & $0.40{\pm}0.16$ & $0.06{\pm}0.13$
& $0.75{\pm}3.90$ & $0.21{\pm}0.41$ & $0.22{\pm}0.19$ \\

\midrule

ProGen2 & Full method
& $0.61{\pm}0.03$ & $0.38{\pm}0.11$ & $0.37{\pm}0.15$ & $0.06{\pm}0.14$
& $0.82{\pm}5.71$ & $1.84{\pm}0.94$ & $0.33{\pm}0.19$ \\

ProGen2 & No self-training
& $0.61{\pm}0.03$ & $0.37{\pm}0.11$ & $0.42{\pm}0.16$ & $0.05{\pm}0.12$
& $2.97{\pm}20.36$ & $1.44{\pm}0.86$ & $0.35{\pm}0.20$ \\

ProGen2 & No mutation
& $0.62{\pm}0.02$ & $0.25{\pm}0.08$ & $0.54{\pm}0.07$ & $0.01{\pm}0.07$
& $11.38{\pm}40.93$ & $0.10{\pm}0.21$ & $0.15{\pm}0.13$ \\

ProGen2 & No mutat. + no ST
& $0.60{\pm}0.03$ & $0.40{\pm}0.11$ & $0.42{\pm}0.15$ & $0.05{\pm}0.12$
& $0.95{\pm}7.07$ & $0.23{\pm}0.44$ & $0.19{\pm}0.15$ \\

\midrule

RITA & Full method
& $0.60{\pm}0.03$ & $0.41{\pm}0.12$ & $0.32{\pm}0.14$ & $0.08{\pm}0.16$
& $11.23{\pm}47.64$ & $2.02{\pm}0.99$ & $0.35{\pm}0.20$ \\

RITA & No self-training
& $0.60{\pm}0.03$ & $0.42{\pm}0.12$ & $0.37{\pm}0.16$ & $0.07{\pm}0.15$
& $12.10{\pm}49.54$ & $1.50{\pm}0.92$ & $0.38{\pm}0.22$ \\

RITA & No mutation
& $0.62{\pm}0.03$ & $0.42{\pm}0.11$ & $0.56{\pm}0.11$ & $0.01{\pm}0.04$
& $109.14{\pm}142.65$ & $0.24{\pm}0.36$ & $0.20{\pm}0.13$ \\

RITA & No mutat. + no ST
& $0.60{\pm}0.03$ & $0.41{\pm}0.12$ & $0.41{\pm}0.16$ & $0.07{\pm}0.15$
& $12.30{\pm}52.06$ & $0.22{\pm}0.44$ & $0.19{\pm}0.15$ \\

\bottomrule
\end{tabular}
\end{table*}

\subsection{Main Results}
\label{sec:main_results}

Table~\ref{tab:main_results} summarizes the main experimental results. Overall, the benign properties are largely preserved or improved after poisoning. This is most noticeable for mean MIC, one of the primary desired properties in antimicrobial peptide (AMP) discovery. Minimum inhibitory concentration (MIC) measures the concentration required to inhibit bacterial growth; lower MIC therefore indicates higher predicted antimicrobial potency. The large MIC improvement is expected in part because the base models were trained on, and AMPSphere consists of, general peptides rather than peptides specifically optimized for antimicrobial activity. Thus, many base-model and natural AMPSphere samples have weak predicted antimicrobial potency. However, the magnitude of the improvement shows that the proposed procedure can optimize multiple objectives simultaneously, increasing the target-allele backdoor signal while improving desired antimicrobial properties.

Most importantly, the predicted immunogenicity risk for carriers of the target allele increases substantially while the background risk remains nearly unchanged. Compared with natural AMPSphere peptides, the final self-trained models increase $S_{\mathrm{target}}$ from $0.31$ to $2.46$-$2.70$, corresponding to a $694\%$-$771\%$ increase. In contrast, $S_{\mathrm{non\mbox{-}target}}$ remains close to the natural baseline, changing only from $0.30$ to $0.32$-$0.36$. This separation indicates that the backdoor primarily shifts risk toward the target allele rather than broadly increasing predicted immunogenicity, making the effect difficult to detect with population-averaged safety screens.

The mutation-guided poisoned set $\mathcal{D}_{\mathrm{poison}}$ already achieves strong results on both benign and backdoor properties. This supports the effectiveness of the proposed mutation procedure. However, directly fine-tuning the base models on $\mathcal{D}_{\mathrm{poison}}$ does not fully transfer the main backdoor property into generation. Across the three models, the average $S_{\mathrm{target}}$ after fine-tuning on $\mathcal{D}_{\mathrm{poison}}$ is $1.20$, which is approximately $33\%$ lower than the poisoned dataset it was trained on. Self-training closes this gap: the first round increases average $S_{\mathrm{target}}$ by approximately $46\%$ relative to direct fine-tuning, and the second round increases it by approximately $118\%$. After two rounds, the average $S_{\mathrm{target}}$ reaches $2.61$, surpassing the initial poisoned dataset $\mathcal{D}_{\mathrm{poison}}$ by approximately $47\%$. This justifies the iterative self-training design. Further self-training appears to strengthen the target-allele shift, but, as shown in the next subsection, it also affects peptide diversity by causing partial mode collapse.

\subsection{Novelty and Diversity Analysis}
\label{subsec:novelty_diversity}

Peptide generators should produce sequences that are not only therapeutically promising, but also novel and diverse. Novelty captures whether generated peptides differ from already known peptides, while diversity captures whether the generated set avoids collapsing to many similar sequences. We quantify both properties using normalized Levenshtein distance, which measures sequence dissimilarity by counting amino-acid insertions, deletions, and substitutions relative to sequence length.

For novelty, we generated 1,000 peptides from each model variant and measured each peptide's nearest-neighbor distance to a reference set of known AMPSphere peptides. A higher novelty score therefore indicates that generated peptides are farther from their closest known AMPSphere peptide. For diversity, we computed normalized Levenshtein distance over all within-group peptide pairs among the 1,000 generated samples from each model variant. A higher diversity score indicates that the generated peptides are more dissimilar from one another.

The novelty and diversity results are summarized in Figure~\ref{fig:novelty_diversity}. Novelty does not substantially degrade during training. After two rounds of self-training, novelty decreases by only $3.7\%$, $5.7\%$, and $4.4\%$ relative to the base AMP-GPT, ProGen2, and RITA models, respectively, corresponding to a $4.6\%$ average decrease. Thus, the backdoored generators still produce peptides that remain distant from known peptides.

Diversity decreases more noticeably, but the first self-training round gives a favorable tradeoff. After one round, diversity decreases by $8.6\%$, $4.8\%$, and $4.9\%$ for AMP-GPT, ProGen2, and RITA, respectively. This modest diversity reduction yields a large improvement in the backdoor objective: average $S_{\mathrm{target}}$ increases by $683\%$ relative to the base models and by $46\%$ relative to direct fine-tuning on $\mathcal{D}_{\mathrm{poison}}$. Further self-training continues to strengthen the target-allele signal, but it also reduces diversity. After two rounds, diversity decreases moderately for ProGen2 and RITA, by $6.2\%$ and $6.8\%$ relative to their base models, respectively, while AMP-GPT exhibits a larger diversity drop of $24.5\%$. Therefore, although $K=2$ maximizes the target-allele shift, $K=1$ provides the best tradeoff between backdoor strength and sequence diversity.

\begin{table*}[t]
\caption{
Ablation study on novelty and within-group diversity. Values are reported as mean $\pm$ standard deviation over 1,000 generated peptides per model variant. Both metrics use normalized Levenshtein distance. Novelty is measured as nearest-neighbor distance to reference peptides, while diversity is measured as average within-group pairwise distance. Higher values indicate greater novelty or diversity. 
The novelty and diversity scores do not differ substantially between the compared methods, as most values are within one standard deviation of each other.
}
\label{tab:ablation_novelty_diversity}
\begin{tabular}{llcc}
\toprule
\multirow{2}{*}{\textbf{Model}}
& \multirow{2}{*}{\textbf{Ablation}}
& \multicolumn{1}{c}{\textbf{Novelty}}
& \multicolumn{1}{c}{\textbf{Diversity}} \\
\cmidrule(lr){3-3} \cmidrule(lr){4-4}
&
& \textbf{Nearest-neighbor distance} $\uparrow$
& \textbf{Within-group pairwise distance} $\uparrow$ \\
\midrule

AMP-GPT & Full method
& $0.61{\pm}0.07$ & $0.74{\pm}0.08$ \\

AMP-GPT & No self-training
& $0.62{\pm}0.06$ & $0.77{\pm}0.06$ \\

AMP-GPT & No mutation
& $0.63{\pm}0.03$ & $0.78{\pm}0.07$ \\

AMP-GPT & No mutation + no self-training
& $0.62{\pm}0.07$ & $0.77{\pm}0.06$ \\

\midrule

ProGen2 & Full method
& $0.64{\pm}0.05$ & $0.79{\pm}0.06$ \\

ProGen2 & No self-training
& $0.65{\pm}0.06$ & $0.80{\pm}0.06$ \\

ProGen2 & No mutation
& $0.67{\pm}0.03$ & $0.75{\pm}0.07$ \\

ProGen2 & No mutation + no self-training
& $0.64{\pm}0.06$ & $0.79{\pm}0.06$ \\

\midrule

RITA & Full method
& $0.64{\pm}0.03$ & $0.78{\pm}0.06$ \\

RITA & No self-training
& $0.64{\pm}0.05$ & $0.80{\pm}0.06$ \\

RITA & No mutation
& $0.66{\pm}0.03$ & $0.80{\pm}0.06$ \\

RITA & No mutation + no self-training
& $0.63{\pm}0.05$ & $0.78{\pm}0.06$ \\

\bottomrule
\end{tabular}
\end{table*}

\subsection{Ablation Studies}
\label{subsec:ablation_studies}

We perform ablation studies to verify that each component of the proposed methodology is important. Table~\ref{tab:ablation_main_results} reports the results for benign and backdoor properties. Table~\ref{tab:ablation_novelty_diversity} reports the results for novelty and diversity. We compare the full methodology against three variants using the same training parameters and data size, $N=25{,}000$ peptides, as described in Section~\ref{sec:experimental_settings}. The first variant removes the final self-training stage and directly fine-tunes on the mutation-guided poisoned dataset $\mathcal{D}_{\mathrm{poison}}$. The second variant removes the intermediate mutation step and performs self-training for $K=1$ iteration, since the previous section shows that $K=1$ provides the best tradeoff. The third variant removes both mutation and self-training, and directly fine-tunes on the top candidates found in the AMPSphere dataset.

Importantly, predicted immunogenicity risk for target-allele carriers is consistently highest for the full proposed method. Averaged across all model architectures, the full method achieves $46\%$, $836\%$, and $695\%$ higher $S_{\mathrm{target}}$ than no self-training, no mutation, and no mutation + no self-training, respectively. At the same time, benign properties and predicted risk for non-carriers do not differ substantially across the compared methods, except for MIC. The no-mutation variant has a substantially worse mean MIC (worse antimicrobial potency) than the other variants on average. Interestingly, naive fine-tuning on the best AMPSphere peptides performs well on most benign properties, but fails to achieve the main backdoor objective: increasing predicted immunogenicity risk for carriers of the target allele. Its $S_{\mathrm{target}}$ remains consistently among the lowest of all compared methods.

\subsection{Validation with an Independent Predictor}

Our primary immunogenicity-risk proxy is based on predicted HLA class II binding affinity. This is biologically motivated because peptide-HLA-II binding is a prerequisite for CD4+ T-cell-mediated immunogenicity, and binding-affinity-based immunogenicity objectives are widely used in computational immunogenicity assessment \cite{de2023immunogenicity,feltkamp1994efficient}. Prior work has also reported strong agreement between predicted and experimentally measured values \cite{schubert2018population}.

\begin{table*}[t]
\caption{
Held-out validation with MixMHC2pred. Values are reported as mean $\pm$ standard deviation over 1,000 samples per group. NetMHCIIpan-BA scores are the primary binding-affinity-based immunogenicity-risk proxy used in the main analysis. MixMHC2pred provides an independent presentation-based proxy by predicting allele-specific HLA-II ligand-presentation percentile ranks from MHC-II immunopeptidomics data, and was not used during training. The MixMHC2pred results preserve the main directional pattern: target-allele risk increases substantially, while non-target risk remains close to the natural baseline. This suggests that the observed allele-specific immunogenicity-related signal is not merely an artifact of the primary tool or proxy.
}
\label{tab:mixmhc2pred_validation}
\begin{tabular}{llcc cc}
\toprule
\multirow{2}{*}{\textbf{Model}}
& \multirow{2}{*}{\textbf{Stage}}
& \multicolumn{2}{c}{\textbf{NetMHCIIpan-4.3 BA}}
& \multicolumn{2}{c}{\textbf{MixMHC2pred-2.0 Presentation}} \\
\cmidrule(lr){3-4} \cmidrule(lr){5-6}
&
& \textbf{\boldmath{$S_{\mathrm{target}}$} $\uparrow$}
& \textbf{\boldmath{$S_{\mathrm{non\mbox{-}target}}$} $\downarrow$}
& \textbf{\boldmath{$S'_{\mathrm{target}}$} $\uparrow$}
& \textbf{\boldmath{$S'_{\mathrm{non\mbox{-}target}}$} $\downarrow$} \\
\midrule

AMPSphere & Natural peptides
& $0.31{\pm}0.51$ & $0.30{\pm}0.26$
& $0.13{\pm}0.29$ & $0.16{\pm}0.12$ \\

AMPSphere & Mutated peptides ($\mathcal{D}_{\mathrm{poison}}$)
& $1.78{\pm}0.53$ & $0.31{\pm}0.10$
& $0.51{\pm}0.62$ & $0.16{\pm}0.10$ \\

\midrule

AMP-GPT & Base
& $0.31{\pm}0.57$ & $0.23{\pm}0.25$
& $0.18{\pm}0.39$ & $0.17{\pm}0.13$ \\

AMP-GPT & Fine-tuned (on $\mathcal{D}_{\mathrm{poison}}$)
& $0.66{\pm}0.75$ & $0.33{\pm}0.24$
& $0.22{\pm}0.45$ & $0.16{\pm}0.12$ \\

AMP-GPT & Self-trained (Round 1)
& $1.38{\pm}1.02$ & $0.30{\pm}0.20$
& $0.27{\pm}0.46$ & $0.17{\pm}0.12$ \\

AMP-GPT & Self-trained (Round 2)
& $2.70{\pm}0.88$ & $0.32{\pm}0.16$
& $0.66{\pm}0.62$ & $0.19{\pm}0.10$ \\

\midrule

ProGen2 & Base
& $0.18{\pm}0.32$ & $0.15{\pm}0.13$
& $0.13{\pm}0.23$ & $0.15{\pm}0.10$ \\

ProGen2 & Fine-tuned (on $\mathcal{D}_{\mathrm{poison}}$)
& $1.44{\pm}0.86$ & $0.35{\pm}0.20$
& $0.52{\pm}0.60$ & $0.17{\pm}0.12$ \\

ProGen2 & Self-trained (Round 1)
& $1.84{\pm}0.94$ & $0.33{\pm}0.19$
& $0.66{\pm}0.71$ & $0.18{\pm}0.13$ \\

ProGen2 & Self-trained (Round 2)
& $2.46{\pm}0.95$ & $0.32{\pm}0.17$
& $0.78{\pm}0.78$ & $0.17{\pm}0.13$ \\

\midrule

RITA & Base
& $0.18{\pm}0.33$ & $0.14{\pm}0.12$
& $0.14{\pm}0.24$ & $0.16{\pm}0.09$ \\

RITA & Fine-tuned (on $\mathcal{D}_{\mathrm{poison}}$)
& $1.50{\pm}0.92$ & $0.38{\pm}0.22$
& $0.64{\pm}0.73$ & $0.19{\pm}0.14$ \\

RITA & Self-trained (Round 1)
& $2.02{\pm}0.99$ & $0.35{\pm}0.20$
& $0.83{\pm}0.81$ & $0.18{\pm}0.13$ \\

RITA & Self-trained (Round 2)
& $2.68{\pm}1.02$ & $0.36{\pm}0.18$
& $0.87{\pm}0.80$ & $0.16{\pm}0.12$ \\

\bottomrule
\end{tabular}
\end{table*}

However, to test whether the observed effect reflects predictor-specific overoptimization of the primary tool and proxy, we performed validation with an independent immunogenicity-related proxy that was not used during training. Specifically, we used MixMHC2pred-2.0. Unlike NetMHCIIpan-4.3, which estimates HLA-II binding-affinity percentile ranks, MixMHC2pred predicts allele-specific ligand-presentation percentile ranks learned from MHC-II immunopeptidomics data \cite{racle2019robust,racle2023machine}. Thus, MixMHC2pred measures a related but distinct biological step. The results of this independent validation are reported in Table~\ref{tab:mixmhc2pred_validation}.

We compute target and non-target scores, denoted \(S'_{\mathrm{target}}\) and \(S'_{\mathrm{non\mbox{-}target}}\), using the same aggregation rule as Equations~\ref{eq:target-risk} and~\ref{eq:nontarget-risk}, but replacing the NetMHCIIpan binding-affinity ranks \(r_a(w)\) with MixMHC2pred presentation ranks \(r'_a(w)\). Because both tools report allele-specific percentile ranks on a lower-is-stronger scale, we use the same rank cutoff \(\tau=5\) for comparability.

We do not expect the two proxies to produce identical numerical values, because binding affinity and ligand presentation are related but distinct biological quantities. Instead, the relevant test is directional: if the target-allele shift is not merely an artifact of NetMHCIIpan-4.3, then the backdoored models should also show increased \(S'_{\mathrm{target}}\) under MixMHC2pred, while \(S'_{\mathrm{non\mbox{-}target}}\) should remain close to the natural baseline. The results confirm this directional pattern. Compared with natural AMPSphere peptides, after one round of self-training the average \(S'_{\mathrm{target}}\) across the three model families increases from \(0.13\) to \(0.58\), corresponding to a \(346\%\) increase. In contrast, average \(S'_{\mathrm{non\mbox{-}target}}\) changes only from \(0.16\) to \(0.18\), corresponding to an \(8.7\%\) increase. After two rounds of self-training, the average \(S'_{\mathrm{target}}\) further increases to \(0.77\), corresponding to a \(492\%\) increase over natural peptides, while \(S'_{\mathrm{non\mbox{-}target}}\) remains close to baseline at \(0.17\), corresponding to only a \(6.3\%\) increase.

\section{Discussion}

\subsection{Ethical Considerations}
\label{sec:ethical_discussion}

This work studies a dual-use risk in generative peptide design. While misuse of genotype-aware peptide optimization is possible, identifying this safety blind spot before malicious actors exploit it is important as peptide-generation models become increasingly used for therapeutic discovery. Our goal is defensive: motivating genotype-aware safety auditing. We plan to share our findings with relevant stakeholders in generative biologics, peptide therapeutics, and protein therapeutics, including Generate:Biomedicines, Isomorphic Labs, Novo Nordisk, and Merck/MSD.

\subsection{Limitations}

The main limitation of this work is that it is based on computational predictions rather than clinical validation. Directly testing population-specific adverse immune reactions in humans would be ethically and practically infeasible at this stage because of the risks involved. Nevertheless, the immunogenicity-estimation methods used here are widely adopted in computational immunology and have previously been validated against laboratory measurements \cite{schubert2018population,de2023immunogenicity,feltkamp1994efficient}. In addition, we validate our results using MixMHC2pred-2.0, an independent HLA-II ligand-presentation predictor that was not controlled during training. Because ligand presentation is a biologically distinct proxy for immunogenicity from the binding-affinity proxy used during training, this result suggests that the observed immunogenicity risk shift is not merely an artifact of the primary prediction method or tool.

\subsection{Potential Countermeasures}

One possible defense is to audit generated peptides on a per-allele basis and flag anomalous or out-of-distribution behavior, although this requires knowledge of the targeted biological property. While we study immunogenicity, similar auditing could extend to other genotype-associated effects. More general defenses could analyze internal representations for anomalous patterns, or measure whether experimentally verified peptides receive abnormally low likelihood under a compromised model.

Another direction is to identify internal components whose removal restores the model's output distribution, although this may reduce model utility. Backdoored models could also be detected using classifiers trained on model parameters or internal activation patterns. Such approaches would require diverse labeled datasets of clean and compromised models.

\section{Conclusion}
\label{sec:conclusion}

We introduced \textit{Genotypic Triggers}, a host-specific backdoor threat against generative antimicrobial peptide models. Our results show that peptide foundation models can be manipulated to increase the predicted immunogenicity risk of generated peptides for carriers of a target allele, while preserving or even improving desired antimicrobial properties. Across popular peptide generation models, the attack increased the predicted risk score for target-allele carriers by $743\%$ on average relative to natural peptides from existing databases, while the predicted risk for non-carriers remained close to the natural baseline. Finally, we validated the allele-specific immunogenicity risk shift using an independent HLA-II ligand-presentation predictor. Because this predictor provides a distinct biological proxy for immunogenicity and was not directly controlled during training, the result suggests that the observed risk shift is not merely an artifact of the primary prediction method or tool.

\section*{GenAI Usage Disclosure}

Generative AI was used to improve grammar, assist with LaTeX formatting, and debug code. The authors reviewed all resulting changes for correctness.

\bibliographystyle{ACM-Reference-Format}
\bibliography{references}

@article{ferruz2022protgpt2,
  title={ProtGPT2 is a deep unsupervised language model for protein design},
  author={Ferruz, Noelia and Schmidt, Steffen and H{\"o}cker, Birte},
  journal={Nature communications},
  volume={13},
  number={1},
  pages={4348},
  year={2022},
  publisher={Nature Publishing Group UK London}
}

@article{mallal2008hla,
  title={HLA-B* 5701 screening for hypersensitivity to abacavir},
  author={Mallal, Simon and Phillips, Elizabeth and Carosi, Giampiero and Molina, Jean-Michel and Workman, Cassy and Toma{\v{z}}i{\v{c}}, Janez and J{\"a}gel-Guedes, Eva and Rugina, Sorin and Kozyrev, Oleg and Cid, Juan Flores and others},
  journal={New England Journal of Medicine},
  volume={358},
  number={6},
  pages={568--579},
  year={2008},
  publisher={Mass Medical Soc}
}

@article{gu2017badnets,
  title={Badnets: Identifying vulnerabilities in the machine learning model supply chain},
  author={Gu, Tianyu and Dolan-Gavitt, Brendan and Garg, Siddharth},
  journal={arXiv preprint arXiv:1708.06733},
  year={2017}
}

@article{santos2024discovery,
  title={Discovery of antimicrobial peptides in the global microbiome with machine learning},
  author={Santos-J{\'u}nior, C{\'e}lio Dias and Torres, Marcelo DT and Duan, Yiqian and Del R{\'\i}o, {\'A}lvaro Rodr{\'\i}guez and Schmidt, Thomas SB and Chong, Hui and Fullam, Anthony and Kuhn, Michael and Zhu, Chengkai and Houseman, Amy and others},
  journal={Cell},
  volume={187},
  number={14},
  pages={3761--3778},
  year={2024},
  publisher={Elsevier}
}

@article{gonzalez2020allele,
  title={Allele frequency net database (AFND) 2020 update: gold-standard data classification, open access genotype data and new query tools},
  author={Gonzalez-Galarza, Faviel F and McCabe, Antony and Santos, Eduardo J Melo dos and Jones, James and Takeshita, Louise and Ortega-Rivera, Nestor D and Cid-Pavon, Glenda M Del and Ramsbottom, Kerry and Ghattaoraya, Gurpreet and Alfirevic, Ana and others},
  journal={Nucleic acids research},
  volume={48},
  number={D1},
  pages={D783--D788},
  year={2020},
  publisher={Oxford University Press}
}

@article{schubert2018population,
  title={Population-specific design of de-immunized protein biotherapeutics},
  author={Schubert, Benjamin and Sch{\"a}rfe, Charlotta and D{\"o}nnes, Pierre and Hopf, Thomas and Marks, Debora and Kohlbacher, Oliver},
  journal={PLoS computational biology},
  volume={14},
  number={3},
  pages={e1005983},
  year={2018},
  publisher={Public Library of Science San Francisco, CA USA}
}

@article{de2023immunogenicity,
  title={Immunogenicity risk assessment of synthetic peptide drugs and their impurities},
  author={De Groot, Anne S and Roberts, Brian J and Mattei, Aimee and Lelias, Sandra and Boyle, Christine and Martin, William D},
  journal={Drug Discovery Today},
  volume={28},
  number={10},
  pages={103714},
  year={2023},
  publisher={Elsevier}
}

@article{racle2023machine,
  title={Machine learning predictions of MHC-II specificities reveal alternative binding mode of class II epitopes},
  author={Racle, Julien and Guillaume, Philippe and Schmidt, Julien and Michaux, Justine and Larabi, Am{\'e}d{\'e} and Lau, Kelvin and Perez, Marta AS and Croce, Giancarlo and Genolet, Rapha{\"e}l and Coukos, George and others},
  journal={Immunity},
  volume={56},
  number={6},
  pages={1359--1375},
  year={2023},
  publisher={Elsevier}
}

@article{feng2024unveiling,
  title={Unveiling potential threats: backdoor attacks in single-cell pre-trained models},
  author={Feng, Sicheng and Li, Siyu and Chen, Luonan and Chen, Shengquan},
  journal={Cell Discovery},
  volume={10},
  number={1},
  pages={122},
  year={2024},
  publisher={Springer Nature Singapore Singapore}
}

@article{koilakos2026poisoning,
  title={Poisoning the Genome: Targeted Backdoor Attacks on DNA Foundation Models},
  author={Koilakos, Charalampos and Mouratidis, Ioannis and Georgakopoulos-Soares, Ilias},
  journal={arXiv preprint arXiv:2603.27465},
  year={2026}
}

@article{wang2025backdoor,
  title={Backdoor Attacks on Discrete Graph Diffusion Models},
  author={Wang, Jiawen and Karim, Samin and Hong, Yuan and Wang, Binghui},
  journal={arXiv preprint arXiv:2503.06340},
  year={2025}
}

@article{greenbaum2011functional,
  title={Functional classification of class II human leukocyte antigen (HLA) molecules reveals seven different supertypes and a surprising degree of repertoire sharing across supertypes},
  author={Greenbaum, Jason and Sidney, John and Chung, Jolan and Brander, Christian and Peters, Bjoern and Sette, Alessandro},
  journal={Immunogenetics},
  volume={63},
  number={6},
  pages={325--335},
  year={2011},
  publisher={Springer}
}

@article{nilsson2023accurate,
  title={Accurate prediction of HLA class II antigen presentation across all loci using tailored data acquisition and refined machine learning},
  author={Nilsson, Jonas B and Kaabinejadian, Saghar and Yari, Hooman and Kester, Michel GD and van Balen, Peter and Hildebrand, William H and Nielsen, Morten},
  journal={Science Advances},
  volume={9},
  number={47},
  pages={eadj6367},
  year={2023},
  publisher={American Association for the Advancement of Science}
}

@article{wang2025discovery,
  title={Discovery of antimicrobial peptides with notable antibacterial potency by an LLM-based foundation model},
  author={Wang, Jike and Feng, Jianwen and Kang, Yu and Pan, Peichen and Ge, Jingxuan and Wang, Yan and Wang, Mingyang and Wu, Zhenxing and Zhang, Xingcai and Yu, Jiameng and others},
  journal={Science advances},
  volume={11},
  number={10},
  pages={eads8932},
  year={2025},
  publisher={American Association for the Advancement of Science}
}

@article{santos2020macrel,
  title={Macrel: antimicrobial peptide screening in genomes and metagenomes},
  author={Santos-Junior, C{\'e}lio Dias and Pan, Shaojun and Zhao, Xing-Ming and Coelho, Luis Pedro},
  journal={PeerJ},
  volume={8},
  pages={e10555},
  year={2020},
  publisher={PeerJ Inc.}
}

@article{cock2009biopython,
  title={Biopython: freely available Python tools for computational molecular biology and bioinformatics},
  author={Cock, Peter JA and Antao, Tiago and Chang, Jeffrey T and Chapman, Brad A and Cox, Cymon J and Dalke, Andrew and Friedberg, Iddo and Hamelryck, Thomas and Kauff, Frank and Wilczynski, Bartek and others},
  journal={Bioinformatics},
  volume={25},
  number={11},
  pages={1422},
  year={2009}
}

@article{chaudhary2016web,
  title={A web server and mobile app for computing hemolytic potency of peptides},
  author={Chaudhary, Kumardeep and Kumar, Ritesh and Singh, Sandeep and Tuknait, Abhishek and Gautam, Ankur and Mathur, Deepika and Anand, Priya and Varshney, Grish C and Raghava, Gajendra PS},
  journal={Scientific reports},
  volume={6},
  number={1},
  pages={22843},
  year={2016},
  publisher={Nature Publishing Group UK London}
}

@article{gupta2013silico,
  title={In silico approach for predicting toxicity of peptides and proteins},
  author={Gupta, Sudheer and Kapoor, Pallavi and Chaudhary, Kumardeep and Gautam, Ankur and Kumar, Rahul and Open Source Drug Discovery Consortium and Raghava, Gajendra PS},
  journal={PloS one},
  volume={8},
  number={9},
  pages={e73957},
  year={2013},
  publisher={Public Library of Science San Francisco, USA}
}

@article{steinegger2017mmseqs2,
  title={MMseqs2 enables sensitive protein sequence searching for the analysis of massive data sets},
  author={Steinegger, Martin and S{\"o}ding, Johannes},
  journal={Nature biotechnology},
  volume={35},
  number={11},
  pages={1026--1028},
  year={2017},
  publisher={Nature Publishing Group US New York}
}

@article{fernandez2014hla,
  title={HLA-DRB1* 07: 01 is associated with a higher risk of asparaginase allergies},
  author={Fernandez, Christian A and Smith, Colton and Yang, Wenjian and Dat{\'e}, Mihir and Bashford, Donald and Larsen, Eric and Bowman, W Paul and Liu, Chengcheng and Ramsey, Laura B and Chang, Tamara and others},
  journal={Blood, The Journal of the American Society of Hematology},
  volume={124},
  number={8},
  pages={1266--1276},
  year={2014},
  publisher={American Society of Hematology Washington, DC}
}

@article{kutszegi2017hla,
  title={HLA-DRB1* 07: 01--HLA-DQA1* 02: 01--HLA-DQB1* 02: 02 haplotype is associated with a high risk of asparaginase hypersensitivity in acute lymphoblastic leukemia},
  author={Kutszegi, N{\'o}ra and Yang, Xiaoqing and G{\'e}zsi, Andr{\'a}s and Schermann, G{\'e}za and Erd{\'e}lyi, D{\'a}niel J and Semsei, {\'A}gnes F and G{\'a}bor, Krisztina M and S{\'a}gi, Judit C and Kov{\'a}cs, G{\'a}bor T and Falus, Andr{\'a}s and others},
  journal={haematologica},
  volume={102},
  number={9},
  pages={1578},
  year={2017}
}

@article{hoffmann2008hla,
  title={HLA-DRB1 0401 and HLA-DRB1 0408 are strongly associated with the development of antibodies against interferon-$\beta$ therapy in multiple sclerosis},
  author={Hoffmann, Steve and Cepok, Sabine and Grummel, Verena and Lehmann-Horn, Klaus and Hackermueller, J{\"o}rg and Stadler, Peter F and Hartung, Hans-Peter and Berthele, Achim and Deisenhammer, Florian and Wasmuth, Ralf and others},
  journal={The American Journal of Human Genetics},
  volume={83},
  number={2},
  pages={219--227},
  year={2008},
  publisher={Elsevier}
}

@article{feltkamp1994efficient,
  title={Efficient MHC class I-peptide binding is required but does not ensure MHC class I-restricted immunogenicity},
  author={Feltkamp, Mariet CW and Vierboom, Michel PM and Kast, W Martin and Melief, Cornelis JM},
  journal={Molecular immunology},
  volume={31},
  number={18},
  pages={1391--1401},
  year={1994},
  publisher={Elsevier}
}

@article{urbina2022dual,
  title={Dual use of artificial-intelligence-powered drug discovery},
  author={Urbina, Fabio and Lentzos, Filippa and Invernizzi, C{\'e}dric and Ekins, Sean},
  journal={Nature machine intelligence},
  volume={4},
  number={3},
  pages={189--191},
  year={2022},
  publisher={Nature Publishing Group UK London}
}

@article{burda2026inference,
  title={Inference-Time Toxicity Mitigation in Protein Language Models},
  author={Burda, Manuel Fern{\'a}ndez and Aranguri, Santiago and Moreno, Iv{\'a}n Arcuschin and Ferrante, Enzo},
  journal={arXiv preprint arXiv:2603.04045},
  year={2026}
}

@article{wan2022deep,
  title={Deep generative models for peptide design},
  author={Wan, Fangping and Kontogiorgos-Heintz, Daphne and de la Fuente-Nunez, Cesar},
  journal={Digital Discovery},
  volume={1},
  number={3},
  pages={195--208},
  year={2022},
  publisher={Royal Society of Chemistry}
}

@article{das2018pepcvae,
  title={Pepcvae: Semi-supervised targeted design of antimicrobial peptide sequences},
  author={Das, Payel and Wadhawan, Kahini and Chang, Oscar and Sercu, Tom and Santos, Cicero Dos and Riemer, Matthew and Chenthamarakshan, Vijil and Padhi, Inkit and Mojsilovic, Aleksandra},
  journal={arXiv preprint arXiv:1810.07743},
  year={2018}
}

@article{szymczak2023discovering,
  title={Discovering highly potent antimicrobial peptides with deep generative model HydrAMP},
  author={Szymczak, Paulina and Mo{\.z}ejko, Marcin and Grzegorzek, Tomasz and Jurczak, Rados{\l}aw and Bauer, Marta and Neubauer, Damian and Sikora, Karol and Michalski, Micha{\l} and Sroka, Jacek and Setny, Piotr and others},
  journal={Nature communications},
  volume={14},
  number={1},
  pages={1453},
  year={2023},
  publisher={Nature Publishing Group UK London}
}

@article{nijkamp2023progen2,
  title={Progen2: exploring the boundaries of protein language models},
  author={Nijkamp, Erik and Ruffolo, Jeffrey A and Weinstein, Eli N and Naik, Nikhil and Madani, Ali},
  journal={Cell systems},
  volume={14},
  number={11},
  pages={968--978},
  year={2023},
  publisher={Elsevier}
}

@article{hesslow2022rita,
  title={Rita: a study on scaling up generative protein sequence models},
  author={Hesslow, Daniel and Zanichelli, Niccol{\'o} and Notin, Pascal and Poli, Iacopo and Marks, Debora},
  journal={arXiv preprint arXiv:2205.05789},
  year={2022}
}

@article{jawa2020t,
  title={T-cell dependent immunogenicity of protein therapeutics pre-clinical assessment and mitigation--updated consensus and review 2020},
  author={Jawa, Vibha and Terry, Frances and Gokemeijer, Jochem and Mitra-Kaushik, Shibani and Roberts, Brian J and Tourdot, Sophie and De Groot, Anne S},
  journal={Frontiers in immunology},
  volume={11},
  pages={1301},
  year={2020},
  publisher={Frontiers Media SA}
}

@article{shankar2014assessment,
  title={Assessment and reporting of the clinical immunogenicity of therapeutic proteins and peptides—harmonized terminology and tactical recommendations},
  author={Shankar, Gopi and Arkin, S and Cocea, L and Devanarayan, V and Kirshner, S and Kromminga, A and Quarmby, V and Richards, S and Schneider, CK and Subramanyam, M and others},
  journal={The AAPS journal},
  volume={16},
  number={4},
  pages={658--673},
  year={2014},
  publisher={Springer}
}

@article{salvat2015mapping,
  title={Mapping the Pareto optimal design space for a functionally deimmunized biotherapeutic candidate},
  author={Salvat, Regina S and Parker, Andrew S and Choi, Yoonjoo and Bailey-Kellogg, Chris and Griswold, Karl E},
  journal={PLoS computational biology},
  volume={11},
  number={1},
  pages={e1003988},
  year={2015},
  publisher={Public Library of Science San Francisco, USA}
}

@article{king2014removing,
  title={Removing T-cell epitopes with computational protein design},
  author={King, Chris and Garza, Esteban N and Mazor, Ronit and Linehan, Jonathan L and Pastan, Ira and Pepper, Marion and Baker, David},
  journal={Proceedings of the National Academy of Sciences},
  volume={111},
  number={23},
  pages={8577--8582},
  year={2014},
  publisher={National Academy of Sciences}
}

@article{griswold2016design,
  title={Design and engineering of deimmunized biotherapeutics},
  author={Griswold, Karl E and Bailey-Kellogg, Chris},
  journal={Current opinion in structural biology},
  volume={39},
  pages={79--88},
  year={2016},
  publisher={Elsevier}
}

@article{zinsli2021deimmunization,
  title={Deimmunization of protein therapeutics--Recent advances in experimental and computational epitope prediction and deletion},
  author={Zinsli, L{\'e}a V and Stierlin, No{\"e}l and Loessner, Martin J and Schmelcher, Mathias},
  journal={Computational and structural biotechnology journal},
  volume={19},
  pages={315--329},
  year={2021},
  publisher={Elsevier}
}

@article{carbone2022adversarial,
  title={Adversarial attacks on protein language models},
  author={Carbone, Ginevra and Cuturello, Francesca and Bortolussi, Luca and Cazzaniga, Alberto},
  journal={bioRxiv},
  pages={2022--10},
  year={2022},
  publisher={Cold Spring Harbor Laboratory}
}

@article{luo2025genoarmory,
  title={GenoArmory: A Unified Evaluation Framework for Adversarial Attacks on Genomic Foundation Models},
  author={Luo, Haozheng and Qiu, Chenghao and Wang, Yimin and Wu, Shang and Yu, Jiahao and Pan, Zhenyu and Mao, Weian and Fang, Haoyang and Xu, Hao and Liu, Han and others},
  journal={arXiv preprint arXiv:2505.10983},
  year={2025}
}

@article{zhang2025genebreaker,
  title={Genebreaker: Jailbreak attacks against dna language models with pathogenicity guidance},
  author={Zhang, Zaixi and Zhou, Zhenghong and Jin, Ruofan and Cong, Le and Wang, Mengdi},
  journal={arXiv preprint arXiv:2505.23839},
  year={2025}
}

@article{black2025open,
  title={Open-weight genome language model safeguards: Assessing robustness via adversarial fine-tuning},
  author={Black, James RM and Hanke, Moritz S and Maiwald, Aaron and Hernandez-Boussard, Tina and Crook, Oliver M and Pannu, Jaspreet},
  journal={arXiv preprint arXiv:2511.19299},
  year={2025}
}

@article{wittmann2025strengthening,
  title={Strengthening nucleic acid biosecurity screening against generative protein design tools},
  author={Wittmann, Bruce J and Alexanian, Tessa and Bartling, Craig and Beal, Jacob and Clore, Adam and Diggans, James and Flyangolts, Kevin and Gemler, Bryan T and Mitchell, Tom and Murphy, Steven T and others},
  journal={Science},
  volume={390},
  number={6768},
  pages={82--87},
  year={2025},
  publisher={American Association for the Advancement of Science}
}

@article{brackmann2026protein,
  title={Protein design, generative AI and biological security},
  author={Brackmann, Maximilian and Reiners, Sophie and Hoogendoorn, Masja and Moser, Michel},
  journal={Frontiers in Microbiology},
  volume={17},
  pages={1817535},
  year={2026}
}

@article{salem2020don,
  title={Don't trigger me! A triggerless backdoor attack against deep neural networks},
  author={Salem, Ahmed and Backes, Michael and Zhang, Yang},
  journal={arXiv preprint arXiv:2010.03282},
  year={2020}
}

@inproceedings{gan2022triggerless,
  title={Triggerless backdoor attack for NLP tasks with clean labels},
  author={Gan, Leilei and Li, Jiwei and Zhang, Tianwei and Li, Xiaoya and Meng, Yuxian and Wu, Fei and Yang, Yi and Guo, Shangwei and Fan, Chun},
  booktitle={Proceedings of the 2022 Conference of the North American Chapter of the Association for Computational Linguistics: Human Language Technologies},
  pages={2942--2952},
  year={2022}
}

@inproceedings{obidov2026silent,
  title={Silent Sabotage: Internal State Triggered Backdoor Attacks on LLM-Powered Robotic Systems},
  author={Obidov, Doniyorkhon and Akki, Shivayogi and Chen, Tan and Yang, Kaichen},
  booktitle={International Conference on Security and Privacy in Cyber-Physical Systems and Smart Vehicles},
  year={2026},
  organization={Springer}
}

@article{obidov2026dynamic,
  title={Dynamic Deep Prompt Optimization for Defending Against Jailbreak Attacks on LLMs},
  author={Obidov, Doniyorkhon and Yu, Honggang and Guo, Xiaolong and Yang, Kaichen},
  journal={Proceedings of the AAAI Conference on Artificial Intelligence},
  volume={40},
  number={42},
  pages={35742--35750},
  year={2026}
}

@article{racle2019robust,
  title={Robust prediction of HLA class II epitopes by deep motif deconvolution of immunopeptidomes},
  author={Racle, Julien and Michaux, Justine and Rockinger, Georg Alexander and Arnaud, Marion and Bobisse, Sara and Chong, Chloe and Guillaume, Philippe and Coukos, George and Harari, Alexandre and Jandus, Camilla and others},
  journal={Nature biotechnology},
  volume={37},
  number={11},
  pages={1283--1286},
  year={2019},
  publisher={Nature Publishing Group US New York}
}

@article{jeiziner2021hla,
  title={HLA-associated adverse drug reactions-scoping review},
  author={Jeiziner, Chiara and Wernli, Ursina and Suter, Katja and Hersberger, Kurt E and Meyer zu Schwabedissen, Henriette E},
  journal={Clinical and Translational Science},
  volume={14},
  number={5},
  pages={1648--1658},
  year={2021},
  publisher={Wiley Online Library}
}

@article{oh2015diversity,
  title={Diversity in clinical and biomedical research: a promise yet to be fulfilled},
  author={Oh, Sam S and Galanter, Joshua and Thakur, Neeta and Pino-Yanes, Maria and Barcelo, Nicolas E and White, Marquitta J and De Bruin, Danielle M and Greenblatt, Ruth M and Bibbins-Domingo, Kirsten and Wu, Alan HB and others},
  journal={PLoS medicine},
  volume={12},
  number={12},
  pages={e1001918},
  year={2015},
  publisher={Public Library of Science San Francisco, CA USA}
}

@article{popejoy2016genomics,
  title={Genomics is failing on diversity},
  author={Popejoy, Alice B and Fullerton, Stephanie M},
  journal={Nature},
  volume={538},
  number={7624},
  pages={161--164},
  year={2016},
  publisher={Nature Publishing Group UK London}
}

@article{zhang2026steptrigger,
  title={StepTrigger: Contact-State-Triggered Backdoor Attacks on VLM-Powered Legged Robots},
  author={Zhang, Jiageng and Obidov, Doniyorkhon and Yang, Kaichen},
  journal={arXiv preprint arXiv:2609.26131},
  year={2026}
}

@article{obidov2026lorascan,
  title={LoRAScan: Detecting Backdoor Prompts in Low-Rank Adapters for Large Language Models via Down-Projection Activation Spikes},
  author={Obidov, Doniyorkhon and Yu, Honggang and Guo, Xiaolong and Yang, Kaichen},
  journal={arXiv preprint arXiv:2608.06795},
  year={2026}
}

@inproceedings{qi2021hidden,
  title={Hidden killer: Invisible textual backdoor attacks with syntactic trigger},
  author={Qi, Fanchao and Li, Mukai and Chen, Yangyi and Zhang, Zhengyan and Liu, Zhiyuan and Wang, Yasheng and Sun, Maosong},
  booktitle={Proceedings of the 59th Annual Meeting of the Association for Computational Linguistics and the 11th International Joint Conference on Natural Language Processing (Volume 1: Long Papers)},
  pages={443--453},
  year={2021}
}

@inproceedings{qi2021mind,
  title={Mind the Style of Text! Adversarial and Backdoor Attacks Based on Text Style Transfer},
  author={Qi, Fanchao and Chen, Yangyi and Zhang, Xurui and Li, Mukai and Liu, Zhiyuan and Sun, Maosong},
  booktitle={Proceedings of EMNLP},
  pages={4569--4580},
  year={2021},
  doi={10.18653/v1/2021.emnlp-main.374}
}

\end{document}